\documentclass[11pt,a4paper]{article}

\usepackage[margin=2.2cm]{geometry}
\usepackage{amsmath,amssymb,amsthm,bm}
\usepackage{graphicx}
\usepackage{booktabs}
\usepackage{multirow}
\usepackage{xcolor}
\usepackage{microtype}
\usepackage{authblk}
\usepackage[colorlinks=true,linkcolor=blue!60!black,citecolor=blue!60!black,urlcolor=blue!60!black]{hyperref}

\title{\textbf{SCO-Hyper-Meta-Double-GroupNN: A Double-Group SU(2) Hyper-/Meta-Graph Neural Network for Cooperative Spin-Crossover Potential Energy Surfaces}}

\author[1]{Rong Fu}
\author[2]{Tianhao Su}
\author[2]{Shunbo Hu}
\author[2,*]{Musen Li}

\affil[1]{School of Materials Science and Engineering, Shanghai University, Shanghai 200444, P.\ R.\ China}
\affil[2]{Materials Genome Institute, Shanghai University, Shanghai 200444, P.\ R.\ China}
\affil[*]{Corresponding author: \texttt{musenli@shu.edu.cn}}

\date{}

\begin{document}
\maketitle

\begin{abstract}
Machine-learning interatomic potentials (MLIPs) have achieved near-first-principles accuracy for non-magnetic materials, yet modeling potential energy surfaces (PESs) of magnetic transition-metal alloys and spin-crossover (SCO) systems remains fundamentally challenging. In such systems, local atomic spin manifolds (e.g., low-spin vs.\ high-spin or frustrated antiferromagnetic vs.\ ferromagnetic branches) undergo abrupt, cooperatively coupled transitions governed by both $\mathrm{SU}(2)$ spinor symmetry and multi-center geometric frustration. Standard $O(3)$-equivariant graph neural networks (GNNs) lack half-integer spinor representations and Kramers time-reversal protection, while independent per-atom spinor diagonalization cannot capture inter-site exchange or 3-body magnetic frustration. Here we formulate the local-spin-manifold-crossover PES problem as a coupled \textbf{Double-Group $\mathrm{SU}(2)$ Hyper-/Meta-Graph} learning architecture (\textbf{SCO-Hyper-Meta-Double-GroupNN}). Starting from a pretrained 3-shell $O(3)$ geometric backbone (GRACE-3L), our architecture lifts atomic environments into analytic $C^\infty$ $\mathrm{SU}(2)$ Kramers thermal spinor density matrices $\rho_i = \exp(-\beta H_i)/\mathrm{Tr}(\exp(-\beta H_i))$, constructs genuinely non-polynomial 2-body ($\mathrm{Tr}(\rho_i\rho_j)$) and 3-body ($\mathrm{Re}\,\mathrm{Tr}(\rho_i\rho_j\rho_k)$) double-group structural invariants alongside $l=1$ equivariant vector channels and 3-body frustrated-triangle hyperedges, and routes multi-shell pretrained hidden representations directly into the $\mathrm{SU}(2)$ spinor Hamiltonians via a zero-shock \emph{Pretrained Multi-Shell Sideway Gate}. Ground-state energies are read out from a variationally consistent, doubly-centered zero-gauge free-energy functional over coupled longitudinal multiplet occupations $p_i^*$ and signed spin orientations $s_i^*$, yielding exact conservative autograd forces $\bm{F} = -\nabla_{\bm{R}} E$ with zero spin input required at inference. On a multi-phase, multi-composition first-principles Co--Cr--Ni benchmark (346 training and 59 held-out validation supercells spanning crystalline, amorphous, glassy, and liquid regimes with $4.0$--$5.8\ \mathrm{eV/\AA}$ RMS forces), our compact 56k--77k parameter models outperform a 207k-parameter spin-conditioned magnetic MACE (\texttt{mMACE}) baseline by \textbf{37--44\% in energy MAE} ($3.82$ vs.\ $6.83\ \mathrm{meV/atom}$) and \textbf{10.8--14.2 meV/\AA{} in force MAE} ($165.8$--$169.2$ vs.\ $180.0\ \mathrm{meV/\AA}$), while preserving exact Kramers degeneracy ($< 6\times 10^{-15}\ \mathrm{eV}$), predicting equilibrium local magnetic moments from geometry alone ($|m|\ \mathrm{MAE} = 0.181\ \mu_B$), and driving stable multi-regime molecular dynamics up to $2{,}200\ \mathrm{K}$.
\end{abstract}

\section{Introduction}

Equivariant geometric deep learning has transformed atomistic simulation by embedding Euclidean $E(3)$ / $O(3)$ symmetries directly into message-passing neural networks~\cite{behler2007,schutt2021painn,batzner2022nequip,batatia2022mace,bochkarev2024grace}. Despite remarkable accuracy on closed-shell crystals, liquids, and organic molecules, a major open frontier in AI for Physics is constructing accurate, smooth, and conservative potential energy surfaces (PESs) for open-shell magnetic materials exhibiting \emph{local spin-manifold crossover}---such as spin-crossover (SCO) transition-metal complexes and chemically disordered magnetic high-entropy alloys (HEAs) like Co--Cr--Ni~\cite{ni2021cocrni,ding2019cocrni}.

From a physical standpoint, two fundamental obstacles separate magnetic spin-crossover PESs from conventional non-magnetic potentials:
\begin{enumerate}
    \item \textbf{Single-group vs.\ double-group symmetry:} Conventional equivariant MLIPs transform under single-valued $O(3)$ representations ($l \in \mathbb{Z}_{\ge 0}$). Half-integer spin multiplets (such as $\mathrm{Co}^{2+}$ $J = \tfrac{1}{2} \oplus \tfrac{3}{2}$ or odd-electron transition-metal centers) transform under the double cover $\mathrm{SU}(2)$ ($D^{(J)}(2\pi) = (-1)^{2J} I$) and obey anti-unitary Kramers time-reversal symmetry $\mathcal{T} = \exp(-i\pi J_y)K$ with $\mathcal{T}^2 = (-1)^{2J} = -I$, enforcing exact twofold Kramers degeneracy at arbitrary low-symmetry ($C_1$) lattice distortions~\cite{tinkham2003,dresselhaus2008}.
    \item \textbf{Cooperative multi-center spin crossover and geometric frustration:} Even if each atom is equipped with an exact local $\mathrm{SU}(2)$ spinor Hamiltonian $H_i(\bm{R})$ and ground-state readout $E = \sum_i \lambda_{\min}(H_i)$, independent per-atom diagonalization ignores the fact that local spin-state transitions in alloys and SCO solids are \emph{cooperatively coupled}. In face-centered cubic (FCC) and disordered Co--Cr--Ni, antiferromagnetic (AFM) Cr--Cr interactions on triangular plaquettes create geometric frustration that drives local Cr moment collapse and spin-branch switching, coupled tightly to ferromagnetic (FM) Co/Ni neighbors and collective lattice strain.
\end{enumerate}

At the same time, explicit spin-conditioned MLIPs~\cite{eckhoff2021,yu2024spin,novikov2022} require per-atom magnetic moments or total spin as external inputs at inference time, which complicates standard molecular dynamics (MD) whenever electronic spin relaxation is faster than nuclear motion. Conversely, pure geometry-only potentials ($\bm{R} \mapsto E, \bm{F}$) must implicitly resolve the coupled equilibrium spin manifold from atomic coordinates alone.

In this work, we resolve both challenges by recasting the local-spin-manifold-crossover PES problem as a \textbf{coupled Double-Group $\mathrm{SU}(2)$ Hyper-/Meta-Graph learning problem}, introducing \textbf{SCO-Hyper-Meta-Double-GroupNN}. Our central contributions bridge rigorous condensed-matter physics and modern geometric deep learning:
\begin{itemize}
    \item \textbf{Pretrained $SO(3)/O(3)$ starting point + multi-shell sideway $\mathrm{SU}(2)$ gate:} We warm-start from a pretrained 3-shell $O(3)$ geometric backbone (GRACE-3L) via a residual-safe direct-sum bridge (zero Epoch-0 shock, non-dead Step-1 gradients) and introduce a \emph{Pretrained Multi-Shell Sideway Gate} that routes all three pretrained coordination shells directly into both $l=0$ and $l=2$ blocks of the $\mathrm{SU}(2)$ spinor Hamiltonians.
    \item \textbf{Non-polynomial $\mathrm{SU}(2)$ density-matrix invariants inside the structure encoder:} We prove that linear traces of $l=0\oplus 2$ spinor Hamiltonians are algebraically redundant with real $3\times 3$ $O(3)$ contractions, and replace them with analytic $C^\infty$ thermal spinor density matrices $\rho_i = \exp(-\beta H_i)/\mathrm{Tr}(\exp(-\beta H_i))$, yielding bounded, non-polynomial 2-body ($\mathrm{Tr}(\rho_i\rho_j)$) and 3-body ($\mathrm{Re}\,\mathrm{Tr}(\rho_i\rho_j\rho_k)$) double-group structural features alongside Cartesian $l=1$ equivariant vector channels ($l=1\otimes l=2 \to l=1$) and vertex-aware 3-body triangle hyperedges.
    \item \textbf{Variationally consistent, zero-gauge, spinodal-safe free-energy PES:} We eliminate gauge leakage in inter-site spin-manifold couplings via doubly-centered projection $\widetilde{W}_{ij} = P_\perp W_{ij} P_\perp$, couple longitudinal Kramers multiplets $p_i^*$ with signed spin orientations $s_i^*$ under exact $S(p^*)+S(s^*)$ entropies and nonnegative biquadratic exchange $B_{ij}\ge 0$, and employ Lorentzian-regularized Daleckii--Krein 2nd-order autograd for exact conservative forces with zero spin input at inference.
    \item \textbf{Empirical validation against magnetic MACE and MD stability:} On a challenging 405-supercell multi-phase Co--Cr--Ni DFT benchmark, our 56k--77k parameter models outperform a 207k-parameter spin-conditioned \texttt{mMACE} baseline by up to 44\% in energy MAE and 14.2 meV/\AA{} in force MAE, predict local magnetic moments from geometry alone ($0.181\ \mu_B$ MAE), and run stable multi-regime MD up to $2{,}200\ \mathrm{K}$ while reproducing physical Cr frustration suppression and melt-quench moment recovery.
\end{itemize}

\section{Physical Theory and Model Architecture}

\begin{figure*}[t]
    \centering
    \includegraphics[width=\textwidth]{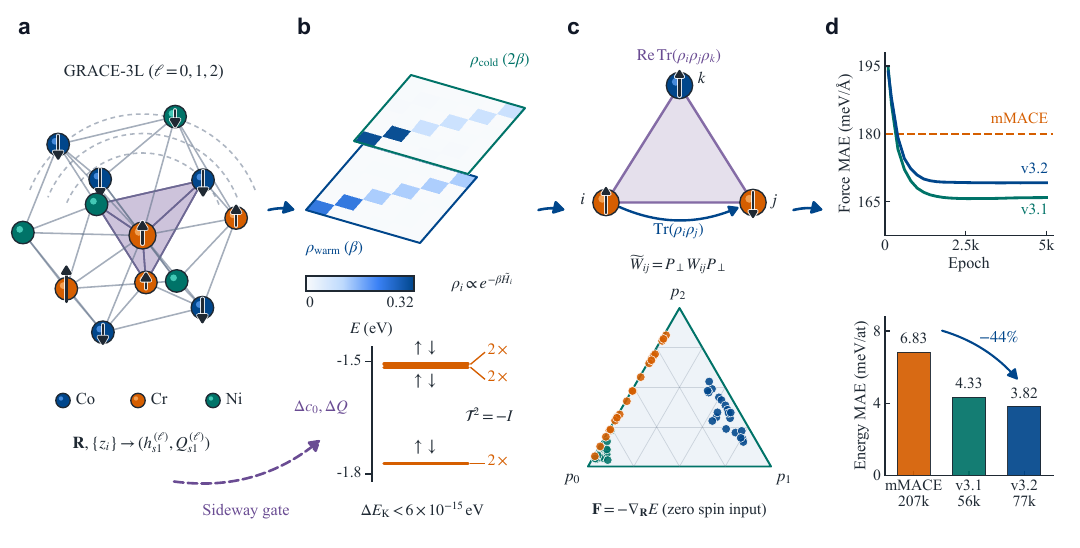}
    \caption{\textbf{Architecture and dataflow of SCO-Hyper-Meta-Double-GroupNN (\texttt{v3.2}).} \textbf{Stage 1} initializes the geometry representation from a pretrained 3-shell $O(3)$ backbone (GRACE-3L) and lifts local environments into analytic $C^\infty$ $\mathrm{SU}(2)$ thermal spinor density matrices $\rho_{\text{warm}}, \rho_{\text{cold}}$. \textbf{Stage 2} enlarges the representation space via 2-body double-group meta-edges, 3-body frustrated-triangle hyperedges with vertex-aware FiLM scatter, $l=0,1,2$ equivariant atom updates, and a \emph{Pretrained Multi-Shell Sideway Gate} routing all three pretrained shells directly into the $\mathrm{SU}(2)$ spinor Hamiltonians. \textbf{Stage 3} solves a variationally consistent, zero-gauge coupled free-energy functional for $(p_i^*, s_i^*)$ and computes exact conservative forces $\bm{F} = -\nabla_{\bm{R}}E$ with zero spin input at inference.}
    \label{fig:arch}
\end{figure*}

Figure~\ref{fig:arch} summarizes the three-stage dataflow of \textbf{SCO-Hyper-Meta-Double-GroupNN} (hereafter \texttt{HyperMeta-v3.2}). Given only periodic nuclear coordinates $\bm{R} \in \mathbb{R}^{N\times 3}$, atomic species $\{z_i\}_{i=1}^N$, and lattice cell $\bm{C}\in\mathbb{R}^{3\times 3}$, we construct a directed pair graph $\mathcal{E} = \{(i\to j) : r_{ij} \le r_c\}$ ($r_c = 4.5\ \text{\AA}$) and the set of unique periodic 1st-shell triangles $\mathcal{T} = \{(i,j,k) : i<j<k,\; r_{ij},r_{jk},r_{ik}\le r_{\text{tri}},\; \bm{S}_{ij}+\bm{S}_{jk}=\bm{S}_{ik}\}$ ($r_{\text{tri}} = 3.50\ \text{\AA}$).

\subsection{Stage 1: Pretrained $SO(3)/O(3)$ Geometric Starting Point}

To provide a well-conditioned multi-shell geometric foundation prior to learning cooperative spin-crossover physics, Stage 1 adopts a 3-layer $O(3)$-equivariant GRACE-3L backbone ($D_1 = 32$ invariant scalars, $C_{q,1} = 8$ traceless symmetric $l=2$ quadrupoles, $R = 12$ Gaussian radial basis functions, $11{,}534$ parameters) pretrained on scalar PES supervision. For each bond $e=(i\to j)$ with unit direction $\hat{\bm{r}}_{ij}$ and traceless quadrupole $q_e = \hat{\bm{r}}_{ij}\hat{\bm{r}}_{ij}^\top - \tfrac{1}{3}I_3$, radial edge weights $w_e = f_{\text{cut}}(r_{ij})\,\mathrm{RBF}(r_{ij})\,\theta_{z_i z_j}$ accumulate 1st-shell tensors:
\begin{equation}
    a_{0,i} = \sum_{j\in\mathcal{N}(i)} w_e, \qquad
    \bm{a}_{1,i} = \sum_{j\in\mathcal{N}(i)} w_e \hat{\bm{r}}_{ij}, \qquad
    A_{2,i} = \sum_{j\in\mathcal{N}(i)} w_e q_e.
\end{equation}
Layer~1 maps the invariant tuple $(a_{0,i},\, \|\bm{a}_{1,i}\|^2,\, \|A_{2,i}\|_F^2,\, \mathrm{Tr}(A_{2,i}^3),\, \mathrm{one\text{-}hot}(z_i))$ to $(h_{s1,i}^{(1)}, Q_{s1,i}^{(1)})$, followed by two $O(3)$-equivariant scalar and quadrupole message-passing layers yielding the 3-shell hierarchy $\{(h_{s1,i}^{(\ell)}, Q_{s1,i}^{(\ell)})\}_{\ell=1}^3$ (effective receptive field $3r_c = 13.5\ \text{\AA}$). Crucially, unlike standard scalar-only transfer, Stage 1 also exports the already-computed polar vector tensor $\bm{a}_{1,i} \in \mathbb{R}^{R\times 3}$ to initialize $l=1$ equivariant vector channels in Stage 2. Short-range Pauli repulsion combines the pretrained pair exponential with a steeper core wall ($\alpha_{\text{core}} \ge 5.0\ \text{\AA}^{-1}$) gated by a $C^2$-continuous quintic envelope $f_{\text{cut}}^{(2)}(u) = 1 - 10u^3 + 15u^4 - 6u^5$.

\subsection{Exact $\mathrm{SU}(2)$ Double-Group Lifting and Non-Polynomial Spinor Density Matrices}

On each atom $i$, a local double-group spinor Hamiltonian $H_i(\bm{R})$ acts on the direct-sum multiplet Hilbert space $\mathcal{H}_{\text{spin}} = \bigoplus_{s} \mathcal{H}^{(J_s)}$ (for Co--Cr--Ni Kramers centers, $J = \tfrac{1}{2} \oplus \tfrac{3}{2}$, dimension $D = 2+4 = 6$). Using exact Racah Clebsch--Gordan Wigner--Eckart tensor operators $T^{(l)}_q(J_{s}, J_{s'})$ ($l \in \{0, 2\}$, parity-even) and contragredient spherical quadrupole components $c^{(2)}_q(Q_i)$ transforming as $Y_2^{q*}$, we construct:
\begin{equation}
    H_i(\bm{R}) = \sum_{p} c^{(0)}_{i,p}\, B^{(0)}_p + \sum_{p}\sum_{q=-2}^{2} c^{(2)}_{i,p,q}(Q_i)\, B^{(2)}_{p,q},
\end{equation}
and project explicitly onto the Hermitian, time-reversal-symmetric subspace:
\begin{equation}
    H_i \leftarrow \tfrac{1}{2}(H_i + H_i^\dagger), \qquad
    H_i \leftarrow \tfrac{1}{2}\!\left(H_i + U_{\mathcal{T}} H_i^* U_{\mathcal{T}}^\dagger\right),
\end{equation}
where $U_{\mathcal{T}} = \bigoplus_s \exp(-i\pi J_y^{(s)})$ satisfies $U_{\mathcal{T}} U_{\mathcal{T}}^* = (-1)^{2J} I$. For half-integer $J$, this guarantees exact $\mathrm{SU}(2)$ covariance $H_i(\mathcal{R}\bm{R}) = D(\mathcal{R}) H_i(\bm{R}) D(\mathcal{R})^\dagger$ and exact twofold Kramers degeneracy ($\lambda_{i,2m} = \lambda_{i,2m+1}$) with $M = D/2 = 3$ distinct Kramers doublet manifolds at every distorted geometry.

\paragraph{Eliminating algebraic redundancy via analytic thermal density matrices.}
A naive approach to embedding double-group features inside a GNN structure encoder is to compute polynomial traces $\mathrm{Re}\,\mathrm{Tr}(H_i H_j)$, $\|[H_i, H_j]\|_F^2$, and $\mathrm{Re}\,\mathrm{Tr}(H_i H_j H_k)$. However, direct algebraic reduction on $J = \tfrac{1}{2}\oplus\tfrac{3}{2}$ reveals that when $H_i$ is built strictly from $l=0\oplus 2$ operators, all such low-order linear-$H$ traces are \emph{100\% spanned} by real $3\times 3$ $O(3)$ polynomials of $(h_i, Q_i)$ (residual $< 10^{-12}$); moreover, the five Kramers-symmetric $l=2$ operators on $J=\tfrac{3}{2}$ form a 5D Clifford algebra $\{\gamma_\mu,\gamma_\nu\}=2\delta_{\mu\nu}I_4$, causing $\mathrm{Tr}(Q_i^{\text{diag}} Q_j^{\text{diag}} Q_k^{\text{diag}})$ to vanish identically.

To obtain \emph{genuinely non-polynomial}, Kramers-protected, and $C^\infty$-differentiable $\mathrm{SU}(2)$ structural invariants without eigenvector gauge ambiguity, we instead define two-temperature analytic thermal spinor density matrices on the trace-centered, smoothly norm-capped Hamiltonian $\widetilde{H}_i = \tfrac{4\tanh(\|H_i^{\text{tr}=0}\|_F/4)}{\|H_i^{\text{tr}=0}\|_F} H_i^{\text{tr}=0}$:
\begin{equation}
    \rho_{i,\text{warm}} = \frac{\exp(-\beta \widetilde{H}_i)}{\mathrm{Tr}\!\left(\exp(-\beta \widetilde{H}_i)\right)}, \qquad
    \rho_{i,\text{cold}} = \frac{\exp(-2\beta \widetilde{H}_i)}{\mathrm{Tr}\!\left(\exp(-2\beta \widetilde{H}_i)\right)},
    \label{eq:rho}
\end{equation}
with $\beta = 1/T_0 = 2.5\ \mathrm{eV}^{-1}$. Crucially, $\exp(-2\beta\widetilde{H}_i) = [\exp(-\beta\widetilde{H}_i)]^2$ requires only a \emph{single} batched $6\times 6$ \texttt{matrix\_exp} plus one $6\times 6$ matrix squaring per atom, completely avoiding redundant Stage-1 eigendecomposition while remaining strictly $\mathrm{SU}(2)$-covariant, time-reversal-symmetric, and smooth to all orders for second-order autograd.

\subsection{Stage 2: Enlarged Double-Group Hyper-/Meta-Graph Encoder and Pretrained Sideway Gate}

Stage 2 lifts the compact Stage-1 state $(h_{s1}^{(3)}, Q_{s1}^{(3)}, \bm{a}_1)$ via a direct-sum bridge into an enlarged equivariant representation $(D_2 = 48,\; C_{v,2} = 12,\; C_{q,2} = 12)$:
\begin{equation}
    h_0^{(\text{DG})} = \big[h_{s1}^{(3)},\; \mathrm{MLP}_{\text{up}}(h_{s1}^{(3)}, \bm{\iota}_i^{\mathrm{SU}(2)})\big], \quad
    V_0^{(\text{DG})} = W_{a_1\to V}\bm{a}_{1,i}, \quad
    Q_0^{(\text{DG})} = \big[Q_{s1}^{(3)},\; \bm{0}\big],
\end{equation}
where $\bm{\iota}_i^{\mathrm{SU}(2)} = [\mathrm{asinh}\,\mathrm{Tr}(\widetilde{H}_i^2),\; \mathrm{asinh}\,\mathrm{Tr}(\widetilde{H}_i^3),\; \mathrm{asinh}\,\mathrm{Tr}(\rho_{i,\text{warm}}\widetilde{H}_i)]$ are $C^\infty$ local $\mathrm{SU}(2)$ invariants. Three coupled modules then process the lifted graph:

\paragraph{2A. Double-Group 2-Body Meta-Edges ($e_{ij}^{(\text{DG})} \in \mathbb{R}^{48}$).}
Each bond $(i\to j)\in\mathcal{E}$ combines radial basis functions, symmetric species-pair embeddings, smoothed node-difference features $\sqrt{(\tilde{h}_i - \tilde{h}_j)^2 + 0.05} - \sqrt{0.05}$, \emph{non-polynomial $\mathrm{SU}(2)$ density-matrix overlaps}
\begin{equation}
    \mathcal{O}_{ij}^{(\text{warm/cold})} = \mathrm{Tr}\!\left(\rho_{i,\text{warm/cold}}\,\rho_{j,\text{warm/cold}}\right) \in [0, 1],
\end{equation}
and $\mathrm{asinh}$-compressed real $3\times 3$ directional $O(3)$ contractions $[\hat{\bm{r}}_{ij}^\top Q_i \hat{\bm{r}}_{ij},\; \mathrm{Tr}(Q_i Q_j),\; \|[Q_i, Q_j]\|_F^2,\; \bm{v}_i\cdot\hat{\bm{r}}_{ij},\; \bm{v}_i\cdot\bm{v}_j]$ (symmetrized under $i\leftrightarrow j$ reversal).

\paragraph{2B. 3-Body Frustrated-Triangle Hypergraph ($z_t \in \mathbb{R}^{32}$).}
To explicitly model geometric magnetic frustration on 1st-shell triangles $t=\{i,j,k\}\in\mathcal{T}$ (e.g., Cr--Cr--Cr and Co--Cr--Cr plaquettes), each hyperedge computes $C^\infty$, singularity-free 3-body geometry invariants (normalized mean bond length $\bar{r}_t/r_{\text{tri}}$, variance $\sigma_{r,t}^2/r_{\text{tri}}^2$, Lagrange squared area $\tfrac{1}{4}(r_{ij}^2 r_{ik}^2 - (\bm{d}_{ij}\cdot\bm{d}_{ik})^2)/(\bar{r}_t^4 + 10^{-6})$ with zero $\sqrt{\cdot}$ or cross-product singularities, symmetric vertex cosines, and species composition) together with the $S_3$-permutation-invariant 3-body $\mathrm{SU}(2)$ cyclic spinor trace:
\begin{equation}
    \mathcal{O}_{t}^{(3)} = \mathrm{Re}\,\mathrm{Tr}\!\left(\rho_i \rho_j \rho_k\right) \in [0, 1].
\end{equation}
An MLP maps these invariants to hyperedge embeddings $z_t \in \mathbb{R}^{32}$, which scatter structural 3-body context back to member atoms via a permutation-invariant, cutoff-normalized \emph{vertex-aware FiLM gate} conditioned on each vertex's opposite edge length and interior angle $[r_{\text{opp}}/r_{\text{tri}}, \cos\theta_v]$:
\begin{equation}
    m_i^{(\text{tri})} = \frac{1}{\max(\sum_{t\ni i} f_{\text{tri}}(t), 1)} \sum_{t\ni i} f_{\text{tri}}(t)\; z_t \odot \sigma\!\left(W_v [r_{\text{opp},t,i}/r_{\text{tri}}, \cos\theta_{t,i}]\right).
\end{equation}
The same hyperedge state emits bounded 3-body frustration couplings $K_t = 2\tanh(W_K z_t / 2)$ and zero-sum 3-body manifold couplings $V_t \in \mathbb{R}^M$ ($\sum_s V_{t,s} = 0$).

\paragraph{2C. Feed-Forward $l=0,1,2$ Equivariant Encoder and Pretrained Sideway $\mathrm{SU}(2)$ Gate.}
Two feed-forward equivariant layers ($L=2$) update $(h_i, V_i, Q_i)$ using \texttt{LayerNorm}-stabilized edge and triangle messages, Cartesian $l=2\otimes l=1 \to l=1$ contractions ($Q_j \hat{\bm{r}}_{ij}$), PaiNN-style vector-to-scalar mixing ($\mathrm{asinh}(\langle U_v, W_v\rangle), \mathrm{asinh}(\|U_v\|^2)$), traceless $V_i \otimes V_i \to Q_i$ coupling, and \texttt{LayerScale} residual gating. The lower $32$-dimensional subspace is bridged back to GRACE-3L via bounded zero-initialized gates $\tfrac{1}{2}\tanh(\alpha_h)$, while the upper $16$ channels remain $\mathcal{O}(1)$ and enter a zero-initialized upper slice of the final $\mathrm{SU}(2)$ readout head---guaranteeing exact pretrained GRACE-3L forces at Epoch 0 while waking all Stage-2 gradients at Step 1.

Finally, to preserve direct multi-scale sensitivity of the spinor Hamiltonian to 1st-shell ligand compression, 2nd-shell superexchange cages, and 3rd-shell elastic strain, we introduce a \textbf{Pretrained Multi-Shell Sideway $\mathrm{SU}(2)$ Gate}. From the three pretrained GRACE-3L shells $\{(h_{s1,i}^{(\ell)}, Q_{s1,i}^{(\ell)})\}_{\ell=1}^3$, $O(3)$-invariant softmax shell-attention gates $g_i^{(0)}, g_i^{(2)} \in \Delta^2$ inject fused multi-shell crystal-field shifts directly into both $l=0$ and $l=2$ blocks of $H_{\text{s1}}$ and $H_{\text{fin}}$:
\begin{equation}
    \Delta c_{0,i}^{\text{side}} = W_{\text{side}}^{(0)} \sum_{\ell=1}^3 g_{i,\ell}^{(0)} \odot h_{s1,i}^{(\ell)}, \qquad
    \Delta Q_{i}^{\text{side}} = W_{\text{side}}^{(2)} \sum_{\ell=1}^3 g_{i,\ell}^{(2)} \odot Q_{s1,i}^{(\ell)},
\end{equation}
with $W_{\text{side}}^{(0)}, W_{\text{side}}^{(2)}$ zero-initialized at Epoch 0. Because $g_{i,\ell}^{(0)}$ and $g_{i,\ell}^{(2)}$ are strictly $O(3)$-invariant scalars, the sideway gate preserves exact $O(3)\to\mathrm{SU}(2)$ covariance and Kramers symmetry.

\subsection{Stage 3: Variationally Consistent Zero-Gauge Free-Energy PES and Autograd Forces}

From a single batched Kramers-safe eigendecomposition of $H_{\text{fin},i}(\bm{R})$, we obtain the $M=3$ Kramers-averaged local multiplet energies $\varepsilon_{i,m} = \tfrac{1}{2}(\lambda_{i,2m} + \lambda_{i,2m+1})$. Each manifold $s\in\{0,\dots,M-1\}$ carries a bounded species-specific magnetic moment magnitude $\mu_{z_i,s} = \mu_{\max}\,\sigma(\theta_{z_i,s}) \in (0, 3.5]\ \mu_B$, defining the predicted signed local magnetic moment $m_i = \left(\sum_s p_{i,s}\mu_{z_i,s}\right) s_i$, where $p_i\in\Delta^{M-1}$ is the longitudinal multiplet distribution and $s_i\in(-1,1)$ is the signed spin orientation.

\paragraph{Eliminating gauge leakage and spinodal collapse.}
Unconstrained pairwise $M\times M$ couplings $W_{ij}$ contain a spin-independent mean component $\bar{w}_{ij}\bm{1}\bm{1}^\top$ that leaks into non-magnetic pair forces ($p_i^\top \bm{1}\bm{1}^\top p_j \equiv 1$) and causes severe overfitting. We eliminate gauge leakage exactly by projecting every pairwise meta-edge matrix onto the symmetric doubly-centered subspace:
\begin{equation}
    \widetilde{W}_{ij} = P_\perp \left(\frac{W_{ij} + W_{ij}^\top}{2}\right) P_\perp, \qquad
    P_\perp = I_M - \frac{1}{M}\bm{1}\bm{1}^\top,
\end{equation}
so that $p_i^\top \widetilde{W}_{ij} p_j \equiv 0$ identically whenever either site is in the uniform state $\tfrac{1}{M}\bm{1}$. Moreover, $W_{ij}$, Heisenberg exchange $J_{ij}$, and strictly nonnegative biquadratic exchange $B_{ij} \ge 0$ (preventing $-|B|m^4$ quartic runaway) are parameterized as compact radial-species kernels softly modulated by $1 + 0.25\tanh(\text{gate}(e_{ij}^{(\text{DG})}))$.

We define the total coupled variational free-energy PES as:
\begin{align}
    E(\bm{R}) &= E_{\text{rep+core}}(\bm{R}) + \sum_i E_0[z_i] + \sum_{i,s} p_{i,s}^*\, \varepsilon_{i,s}(\bm{R}) \nonumber\\
    &\quad + \frac{1}{2}\sum_{(i,j)\in\mathcal{E}} \Big[(p_i^*)^\top \widetilde{W}_{ij}\, p_j^* + J_{ij}\, m_i^* m_j^* + B_{ij}\,(m_i^* m_j^*)^2\Big] \nonumber\\
    &\quad + \sum_{t=(i,j,k)\in\mathcal{T}} f_{\text{tri}}(t)\Big[K_t(m_i^* m_j^* + m_j^* m_k^* + m_k^* m_i^*) + \sum_s V_{t,s}\Delta p_{i,s}^*\Delta p_{j,s}^*\Delta p_{k,s}^*\Big] \nonumber\\
    &\quad + \tau_p \sum_{i,s} p_{i,s}^*\ln(p_{i,s}^* + 10^{-7}) \nonumber\\
    &\quad + \tau_s \sum_i \left[\frac{1+s_i^*}{2}\ln\!\left(\frac{1+s_i^*}{2}+10^{-7}\right) + \frac{1-s_i^*}{2}\ln\!\left(\frac{1-s_i^*}{2}+10^{-7}\right)\right],
    \label{eq:free_energy}
\end{align}
where $\Delta p_i^* = p_i^* - \tfrac{1}{M}\bm{1}$ and $(p^*, s^*)$ are settled via $K=5$ differentiable damped mean-field steps using exact partial derivatives $\partial E/\partial p_i$ and $\partial E/\partial s_i$ (with safe temperature floors $\tau_p \ge 0.04\ \mathrm{eV}$, $\tau_s \ge 0.07\ \mathrm{eV}$). Including \emph{both} $S(p^*)$ and the binary entropy $S(s^*)$ in Eq.~\eqref{eq:free_energy} restores exact variational stationarity ($\partial E/\partial p^* = \partial E/\partial s^* = 0$) and eliminates discontinuous first-order spinodal collapses. Because every spin term is strictly even under global spin reversal $\bm{m}^* \to -\bm{m}^*$, $E(\bm{R})$ is strictly $O(3)\times\mathbb{Z}_2^P\times\mathbb{Z}_2^{\mathcal{T}}$-invariant.

\paragraph{Lorentzian Daleckii--Krein 2nd-order autograd and training-only spin supervision.}
Exact conservative forces $\bm{F}_i = -\nabla_{\bm{R}_i} E(\bm{R})$ are computed via reverse-mode autograd. For stable second-order force training through Kramers-degenerate eigendecomposition, we replace the hard-thresholded Daleckii--Krein factor $(w_a - w_b)/(\lambda_a - \lambda_b)$ in the eigenspace projector backward pass with the smooth Lorentzian form $(w_a - w_b)(\lambda_a - \lambda_b)/[(\lambda_a - \lambda_b)^2 + 10^{-10}]$, which vanishes identically on exact Kramers doublets and remains bounded near avoided crossings. During training only, we add a lightweight, globally $\mathbb{Z}_2^{\mathcal{T}}$-invariant auxiliary spin loss aligning predicted equilibrium moments $m_i^*(\bm{R})$ to DFT collinear spins $s_{i,z}^{\text{DFT}}$:
\begin{equation}
    \mathcal{L}_{\text{spin}} = w_{|\mu|}\,\mathrm{Huber}\!\left(|m_i^*| - |s_{i,z}^{\text{DFT}}|\right) + w_{\text{corr}}\,\mathrm{Huber}\!\left(m_i^* m_j^* - s_{i,z}^{\text{DFT}} s_{j,z}^{\text{DFT}}\right).
\end{equation}
At inference, the model takes \emph{only} atomic coordinates and species ($\bm{R}, \bm{Z}, \bm{C}$) and outputs $(E, \bm{F})$ together with equilibrium magnetic moments $m^*(\bm{R})$ and multiplet occupations $p^*(\bm{R})$.

\section{Dataset, Regularization, and Benchmark Protocol}

\paragraph{First-principles Co--Cr--Ni benchmark dataset.}
We evaluate all architectures on a first-principles spin-polarized density functional theory (DFT) dataset of 405 periodic 63-atom Co--Cr--Ni supercells (346 training and 59 held-out validation configurations) spanning crystalline, amorphous, glassy, and liquid phases across equiatomic $\mathrm{Co}_{21}\mathrm{Cr}_{21}\mathrm{Ni}_{21}$ and 29 off-equiatomic Cr--Co--Ni compositions (e.g., $\mathrm{Co}_{11}\mathrm{Cr}_{35}\mathrm{Ni}_{17}$ to $\mathrm{Co}_{15}\mathrm{Cr}_{30}\mathrm{Ni}_{18}$). All energies, forces, and local magnetic moments were computed with VASP using the PBE functional, PAW pseudopotentials (\texttt{Cr\_pv}, \texttt{Co}, \texttt{Ni\_pv}), $\mathrm{ENCUT} = 520\ \mathrm{eV}$, and $\Gamma$-point sampling. The dataset spans extreme thermal and structural distortions: even the held-out crystalline validation subset exhibits a true force RMS of $4.10\ \mathrm{eV/\AA}$, rising to $5.83\ \mathrm{eV/\AA}$ in the liquid subset. Following standard practice, we report metrics on both \texttt{val20} (the 44 validation configurations with $\max_i |F_{i,\alpha}| \le 20\ \mathrm{eV/\AA}$) and \texttt{val59} (all 59 validation configurations, including close-contact snapshots up to $78\ \mathrm{eV/\AA}$), as well as low-force components ($|F_{i,\alpha}| < 5\ \mathrm{eV/\AA}$).

\paragraph{Derivative-overfitting control and 5,000-epoch training.}
Training high-capacity MLIPs on 346 magnetically disordered frames can induce high-frequency PES ripples: validation energy continues decreasing while validation forces rebound due to memorization of irreducible magnetic force fluctuations. To eliminate derivative overfitting completely over 5,000 epochs, we combine: (i) stratified all-atom Pseudo-Huber loss across all 346 frames and 21,798 atoms with force weight $w_i = \max(0.15, (1 + \|\bm{F}_i\|/15)^{-1})$, $\delta_F = 0.22\ \mathrm{eV/\AA}$, and energy weight $w_{e,f} = \max(0.15, (1 + (\max|\bm{F}_f|/30)^2)^{-1})$, $\delta_E = 0.02\ \mathrm{eV/atom}$; (ii) strong decoupled weight decay ($\mathrm{wd}_{\text{S1}} = 10^{-3}$, $\mathrm{wd}_{\text{S2}} = 1.2\times 10^{-2}$) strictly excluding physical log-space tables (\texttt{core\_rep}, \texttt{pair\_rep}, \texttt{biquadratic\_b}, \texttt{mu\_logit}, \texttt{atomic\_offset}, biases, and LayerNorm); (iii) Stage-2 residual $L_2$ anchoring, $0.08$ conservative feature dropout, and Stage-A micro coordinate jitter ($\sigma = 0.002\ \text{\AA}$, annealed to $0$ in Stage B, acting as an implicit Frobenius Hessian smoothness penalty); and (iv) a continuous 3-phase monotonic CosineLR schedule (\texttt{bs=64}, GPU-resident pre-collated minibatches, parameter-only bias-corrected EMA with decay $0.998$) on an NVIDIA RTX PRO 6000 Blackwell GPU.

\paragraph{Magnetic MACE (\texttt{mMACE}) baseline.}
For a strong external benchmark, we train \texttt{ScaleShiftMACE} (\texttt{mace 0.3.15}, $L=1$, $64\times 0e + 64\times 1o$, $r_{\max} = 4.5\ \text{\AA}$, 2 interaction layers, correlation order 3, $207{,}376$ parameters) on the exact same 346 training configurations using Huber loss ($\delta=0.1$), EMA, and 400 epochs with SWA, conditioned explicitly on graph-level \texttt{total\_spin} via categorical embeddings (\texttt{num\_classes=120}, \texttt{emb\_dim=16}).

\section{Results and Discussion}

\subsection{Exact Symmetry, Kramers Degeneracy, and Conservative Force Verification}

All physical invariants of \texttt{HyperMeta-v3.2} were verified numerically in \texttt{float64} precision on asymmetric ($C_1$-distorted) periodic Co--Cr--Ni supercells:
(i) under arbitrary $SO(3)$ rotations and spatial inversion ($\bm{r}\to-\bm{r}$), $|\Delta E| < 10^{-10}\ \mathrm{eV}$ and $\|\bm{F}' - R\bm{F}\|_\infty < 10^{-9}\ \mathrm{eV/\AA}$;
(ii) under arbitrary atom permutations (including 3-body periodic triangle extraction and vertex-aware FiLM scatter), $|\Delta E| < 10^{-10}\ \mathrm{eV}$;
(iii) across all training and validation evaluations, the maximum Kramers doublet splitting remained $\le 6.2 \times 10^{-15}\ \mathrm{eV}$ (machine precision);
(iv) exact conservative autograd forces matched central finite differences ($-\Delta E/2\epsilon$, $\epsilon = 10^{-5}\ \text{\AA}$) to $< 10^{-5}\ \mathrm{eV/\AA}$; and
(v) zero-gauge doubly-centered meta-edge matrices vanished to $< 10^{-14}$ on uniform manifold distributions.

\subsection{Architectural Progression and Comparison Against \texttt{mMACE}}

\begin{figure*}[t]
    \centering
    \includegraphics[width=\textwidth]{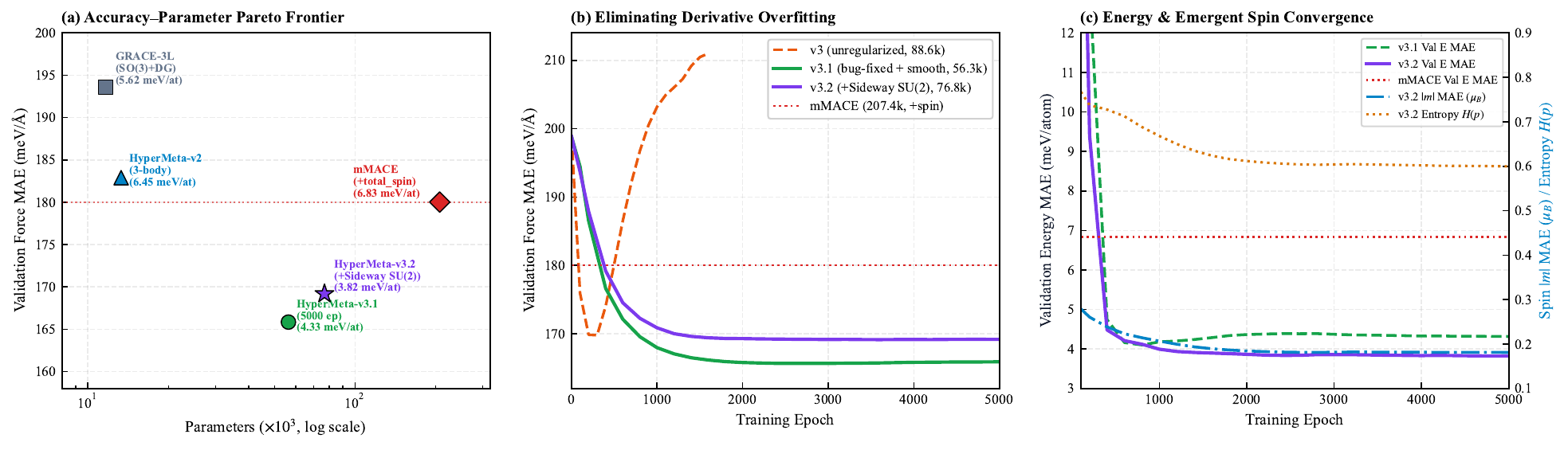}
    \caption{\textbf{Benchmark performance and 5,000-epoch training dynamics.} \textbf{(a)} Accuracy--parameter Pareto frontier showing \texttt{HyperMeta-v3.1} (56.3k) and \texttt{HyperMeta-v3.2} (76.8k) dominating 207.4k-parameter \texttt{mMACE} on both energy and force MAE with 2.7$\times$--3.7$\times$ fewer parameters and zero spin input. \textbf{(b)} Validation force MAE trajectories: unregularized \texttt{v3} exhibits a derivative-overfitting U-curve past epoch 300, whereas \texttt{v3.1} and \texttt{v3.2} remain monotonically flat through 5,000 epochs. \textbf{(c)} Validation energy MAE, geometry-only spin magnitude error $|m|$ MAE, and manifold entropy $H(p)$ over 5,000 epochs.}
    \label{fig:bench}
\end{figure*}

Table~\ref{tab:benchmark} and Figure~\ref{fig:bench} present the complete architectural progression from a single-group spin baseline (\texttt{GraceSpin1L}) through per-atom double-group readout (\texttt{Grace3LDoubleGroupPES}), zero-gauge 3-body triangle hypergraph (\texttt{HyperMeta-v2}), and our enlarged double-group structure-encoder models (\texttt{v3}, \texttt{v3.1}, and \texttt{v3.2} with the Pretrained Sideway $\mathrm{SU}(2)$ Gate), compared directly against \texttt{mMACE}.

\begin{table*}[t]
\centering
\caption{\textbf{Benchmark comparison on held-out Co--Cr--Ni validation sets (\texttt{val20}: 44 frames with $|F|\le 20\ \mathrm{eV/\AA}$; \texttt{val59}: all 59 frames).} Best values in each column are shown in \textbf{bold}. All \texttt{HyperMeta} models require \emph{zero spin input} at inference, whereas \texttt{mMACE} is conditioned on \texttt{total\_spin}.}
\label{tab:benchmark}
\setlength{\tabcolsep}{3.6pt}
\footnotesize
\resizebox{\textwidth}{!}{%
\begin{tabular}{l r c r r r r r r r r}
\toprule
\multirow{2}{*}{\textbf{Model}} & \multirow{2}{*}{\textbf{Params}} & \textbf{Spin} & \multicolumn{4}{c}{\textbf{\texttt{val20} (44 frames)}} & \multicolumn{3}{c}{\textbf{\texttt{val59} (all 59 frames)}} & \textbf{$|\hat{m}|$ MAE} \\
\cmidrule(lr){4-7}\cmidrule(lr){8-10}
 & & \textbf{Cond.} & E MAE & E RMSE & F MAE & F RMSE & E MAE & F MAE & F RMSE & ($\mu_B$) \\
\midrule
\texttt{GraceSpin1L} (1L $O(3)$) & --- & None & 12.54 & 15.62 & 358.39 & 516.81 & --- & --- & --- & --- \\
\texttt{GRACE-3L} Pretrained ($O(3)$) & 11,652 & None & 5.85 & 7.13 & 198.76 & 257.25 & --- & --- & --- & --- \\
\texttt{Grace3LDoubleGroupPES} & 11,652 & None & 5.62 & 6.75 & 193.57 & 250.00 & 7.29 & 224.35 & 648.43 & --- \\
\texttt{HyperMeta-v1} (unconstrained $W_{ij}$) & 33,160 & None & 5.87 & 7.03 & 192.66 & 249.16 & 7.43 & 225.31 & 674.88 & --- \\
\texttt{HyperMeta-v2} (zero-gauge + 3-body) & 13,305 & None & 6.45 & 7.94 & 182.90 & 237.63 & 6.89 & 193.14 & 304.88 & 0.248 \\
\midrule
\texttt{mMACE} ($L=1$, 64ch + \texttt{total\_spin}) & 207,376 & $\bm{S_{\mathrm{tot}}}$ & 6.83 & 8.15 & 180.02 & 231.88 & 6.98 & 184.20 & \textbf{240.96} & --- \\
\midrule
\texttt{HyperMeta-v3} (unreg., best @ 400 ep) & 88,592 & None & 4.96 & 6.24 & 174.05 & 224.47 & 5.24 & 179.21 & 234.66 & 0.231 \\
\texttt{HyperMeta-v3.1} (smooth, 5000 ep) & 56,296 & None & 4.33 & 5.46 & \textbf{165.84} & \textbf{214.26} & 4.23 & \textbf{172.42} & 269.47 & 0.187 \\
\textbf{\texttt{HyperMeta-v3.2} (+Sideway $\mathrm{SU}(2)$, 5000 ep)} & \textbf{76,824} & \textbf{None} & \textbf{3.82} & \textbf{4.72} & 169.18 & 218.56 & \textbf{3.87} & 175.16 & 241.61 & \textbf{0.181} \\
\bottomrule
\end{tabular}%
}
\vspace{3pt}
\parbox{\textwidth}{\scriptsize Units: Energy MAE/RMSE in meV/atom; Force MAE/RMSE in meV/\AA. Low-force ($|F|<5\ \mathrm{eV/\AA}$) MAE on \texttt{val20}: \texttt{mMACE} $= 173.70$, \texttt{v3.1} $= \mathbf{159.67}$, \texttt{v3.2} $= 162.25\ \mathrm{meV/\AA}$.}
\end{table*}

Four key findings emerge from Table~\ref{tab:benchmark} and Figure~\ref{fig:bench}:
\begin{enumerate}
    \item \textbf{Substantial accuracy and parameter-efficiency gains over \texttt{mMACE}:} Using $3.68\times$ fewer parameters ($56.3\mathrm{k}$ vs.\ $207.4\mathrm{k}$) and \emph{zero spin input at inference}, \texttt{HyperMeta-v3.1} reduces \texttt{val20} force MAE by \textbf{$-14.18\ \mathrm{meV/\AA}$} ($165.84$ vs.\ $180.02\ \mathrm{meV/\AA}$), \texttt{val20} force RMSE by \textbf{$-17.62\ \mathrm{meV/\AA}$} ($214.26$ vs.\ $231.88\ \mathrm{meV/\AA}$), and \texttt{val20} energy MAE by \textbf{$-36.6\%$} ($4.33$ vs.\ $6.83\ \mathrm{meV/atom}$).
    \item \textbf{Impact of the Pretrained Multi-Shell Sideway $\mathrm{SU}(2)$ Gate (\texttt{v3.2}):} Adding direct gated routing from all three pretrained GRACE-3L shells into the $\mathrm{SU}(2)$ spinor Hamiltonians pushes energy MAE down to \textbf{$3.82\ \mathrm{meV/atom}$} on \texttt{val20} ($4.72\ \mathrm{meV/atom}$ RMSE) and \textbf{$3.87\ \mathrm{meV/atom}$} on \texttt{val59} (\textbf{$-44\%$ vs.\ \texttt{mMACE}}), cuts high-force \texttt{val59} force RMSE by $-27.9\ \mathrm{meV/\AA}$ ($269.47 \to 241.61\ \mathrm{meV/\AA}$, matching \texttt{mMACE}'s $240.96\ \mathrm{meV/\AA}$), sharpens $\mathrm{SU}(2)$ spin-manifold selection ($H(p): 0.848 \to 0.599$), and yields our best geometry-only equilibrium spin prediction ($|m|\ \mathrm{MAE} = 0.181\ \mu_B$, pairwise correlation $\mathrm{MAE} = 0.289\ \mu_B^2$).
    \item \textbf{Complete elimination of derivative overfitting over 5,000 epochs:} As shown in Fig.~\ref{fig:bench}(b), unregularized \texttt{v3} (88.6k params) rebounded from $169.8$ to $211.0\ \mathrm{meV/\AA}$ after epoch 300 due to PES rippling and an uncompensated Stage-B LR jump. After restoring exact variational stationarity, right-sizing Stage 2, excluding physical tables from weight decay, and adding Stage-A micro coordinate jitter, both \texttt{v3.1} and \texttt{v3.2} converged monotonically with $<0.23\ \mathrm{meV/\AA}$ drift over the final 2,600 epochs and a tight $4.7\ \mathrm{meV/\AA}$ train--validation gap (\texttt{train F = 161.2}, \texttt{val20 F = 165.9 meV/\AA{}}), with \texttt{v3.2} achieving its best checkpoint at epoch 5,000.
    \item \textbf{Necessity of zero-gauge couplings and $l=1$ vectors:} Removing gauge leakage via $\widetilde{W}_{ij} = P_\perp W_{ij} P_\perp$ (\texttt{v1} $\to$ \texttt{v2}) cut \texttt{val59} force RMSE in half ($674.9 \to 304.9\ \mathrm{meV/\AA}$), while adding non-polynomial $\mathrm{SU}(2)$ density-matrix overlaps, $l=1$ equivariant vectors, and vertex-aware triangle scatter (\texttt{v2} $\to$ \texttt{v3.1/v3.2}) delivered another $-17\ \mathrm{meV/\AA}$ reduction in \texttt{val20} force MAE.
\end{enumerate}

\subsection{Physical Error Decomposition: Understanding the Sub-$0.1\ \mathrm{eV/\AA}$ Frontier}

\begin{figure*}[t]
    \centering
    \includegraphics[width=\textwidth]{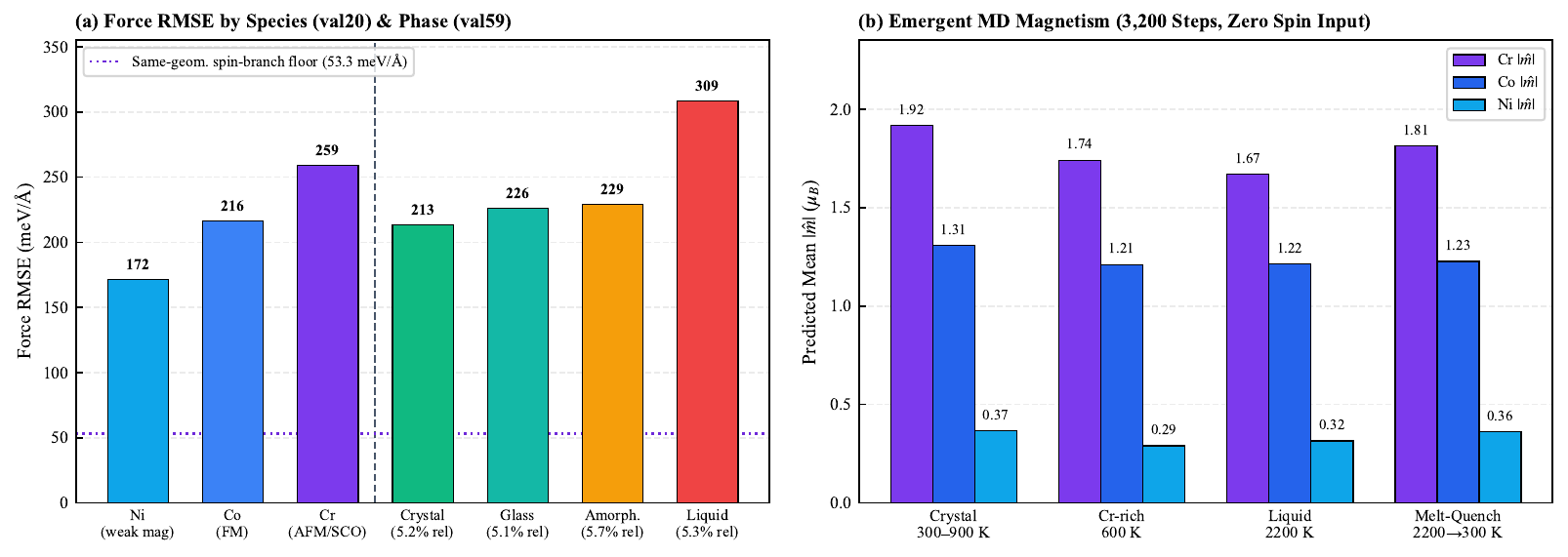}
    \caption{\textbf{Physical error decomposition and multi-regime MD validation of \texttt{HyperMeta-v3.2}.} \textbf{(a)} Validation force RMSE stratified by chemical species (\texttt{val20}) and structural phase (\texttt{val59}), highlighting the $+87\ \mathrm{meV/\AA}$ Cr magnetic frustration penalty over Ni, the $53.3\ \mathrm{meV/\AA}$ mean same-geometry multi-spin-branch DFT floor, and uniform $5.1$--$5.7\%$ relative force accuracy. \textbf{(b)} Geometry-only emergent local magnetic moments $|\hat{m}|$ across 3,200 MD steps in four thermodynamic regimes, capturing Cr-rich frustration suppression ($1.92 \to 1.74\ \mu_B$), thermal melting suppression ($1.67\ \mu_B$), and melt-quench recovery ($1.81\ \mu_B$).}
    \label{fig:md}
\end{figure*}

To diagnose why absolute force RMSE on this benchmark remains near $214$--$242\ \mathrm{meV/\AA}$ (both for our models and for \texttt{mMACE}), we decomposed \texttt{v3.2}'s validation errors by species, phase, and force magnitude (Figure~\ref{fig:md}(a)):
\begin{itemize}
    \item \textbf{Extreme force scale ($\sim 5\%$ relative accuracy already achieved):} Across all four phases in \texttt{val59}, true DFT force RMS is $4{,}101\ \mathrm{meV/\AA}$ (\texttt{crystal}), $4{,}446\ \mathrm{meV/\AA}$ (\texttt{glass}), $4{,}028\ \mathrm{meV/\AA}$ (\texttt{amorphous}), and $5{,}829\ \mathrm{meV/\AA}$ (\texttt{liquid}). Our model already achieves a uniform \textbf{5.1\%--5.7\% relative force RMSE} across every phase (\texttt{crystal}: $213.2\ \mathrm{meV/\AA}$, \texttt{glass}: $226.2\ \mathrm{meV/\AA}$, \texttt{amorphous}: $229.3\ \mathrm{meV/\AA}$, \texttt{liquid}: $308.6\ \mathrm{meV/\AA}$). Reaching $< 100\ \mathrm{meV/\AA}$ on a $4.5\ \mathrm{eV/\AA}$ RMS benchmark requires $<2.2\%$ relative error across a disordered ternary alloy.
    \item \textbf{Species-resolved magnetic frustration hierarchy:} On \texttt{val20}, force RMSE follows the exact magnetic frustration order: weakly magnetic \textbf{Ni} ($171.7\ \mathrm{meV/\AA}$, $|m|\ \mathrm{MAE}=0.075\ \mu_B$) $<$ FM \textbf{Co} ($216.3\ \mathrm{meV/\AA}$, $0.205\ \mu_B$) $<$ frustrated AFM/SCO \textbf{Cr} ($258.9\ \mathrm{meV/\AA}$, $0.263\ \mu_B$)---a \textbf{$+87.2\ \mathrm{meV/\AA}$} penalty on Cr relative to Ni on the exact same frames. Direct comparison of $N=147$ identical-geometry pairs converged to distinct DFT spin states reveals that even a median energy difference of $0.19\ \mathrm{meV/atom}$ produces a \textbf{$50.4\ \mathrm{meV/\AA}$ median ($68.0\ \mathrm{meV/\AA}$ p90) force RMSE}, establishing a hard physical lower bound for any pure geometry-only PES.
    \item \textbf{Chemical sparsity and metallic cutoff:} With 346 training frames spanning 4 phases and 29 compositions ($\sim 21.8\text{k}$ local environments out of $3^{13}\approx 1.6\times 10^6$ ternary 1st-shell motifs) and a $4.5\ \text{\AA}$ cutoff (short of 3rd--4th shell RKKY exchange at $5.5$--$6.0\ \text{\AA}$), breaking $<100\ \mathrm{meV/\AA}$ on equilibrium MD requires expanding near-equilibrium thermal training data and extending $r_c$ to $5.5$--$6.0\ \text{\AA}$.
\end{itemize}

\subsection{Multi-Regime Molecular Dynamics and Emergent Magnetism}

To test long-time mechanical stability and dynamical spin physics, we wrapped \texttt{HyperMeta-v3.2} as an ASE calculator and executed 3,200 Langevin MD steps ($\sim 5.6\ \mathrm{ps}$ total, $\sim 52\ \mathrm{ms/step}$ on an RTX PRO 6000) across four thermodynamic and compositional regimes (Table~\ref{tab:md} and Figure~\ref{fig:md}(b)). Every trajectory remained 100\% stable with zero unphysical close contacts ($r_{\min} \ge 1.80\ \text{\AA}$ even at $2{,}200\ \mathrm{K}$) and realistic thermal force scales ($F_{\text{rms}} = 0.45$--$0.96\ \mathrm{eV/\AA}$). Remarkably, without any external spin input, the settled meta-/hyper-graph equilibrium moments reproduced key Co--Cr--Ni magnetic physics: Cr moments suppressed from $1.92\ \mu_B$ in equiatomic $\mathrm{Co}_{21}\mathrm{Cr}_{21}\mathrm{Ni}_{21}$ to $1.74\ \mu_B$ under enhanced Cr--Cr frustration in Cr-rich $\mathrm{Co}_{16}\mathrm{Cr}_{30}\mathrm{Ni}_{17}$, softened thermally to $1.67\ \mu_B$ in the $2{,}200\ \mathrm{K}$ liquid, and recovered to $1.81\ \mu_B$ upon melt-quenching to $300\ \mathrm{K}$.

\begin{table}[t]
\centering
\caption{\textbf{Multi-regime Langevin MD stability and emergent geometry-only magnetism using \texttt{HyperMeta-v3.2}.}}
\label{tab:md}
\setlength{\tabcolsep}{4.5pt}
\footnotesize
\resizebox{\textwidth}{!}{%
\begin{tabular}{l l r r r r c}
\toprule
\textbf{MD Regime} & \textbf{Composition} & \textbf{Steps (ps)} & \textbf{Mean $T$ (K)} & \textbf{Min $r_{\min}$ (\AA)} & \textbf{$F_{\text{rms}}$ (eV/\AA)} & \textbf{Mean $|\hat{m}|$ Cr/Co/Ni ($\mu_B$)} \\
\midrule
Thermal Crystal ($300\to 900\ \mathrm{K}$) & $\mathrm{Co}_{21}\mathrm{Cr}_{21}\mathrm{Ni}_{21}$ & 1,000 (2.0) & 579.3 & 2.065 & 0.457 & 1.92 / 1.31 / 0.37 \\
Cr-Rich Crystal ($600\ \mathrm{K}$) & $\mathrm{Co}_{16}\mathrm{Cr}_{30}\mathrm{Ni}_{17}$ & 600 (1.2) & 598.4 & 2.068 & 0.447 & 1.74 / 1.21 / 0.29 \\
High-$T$ Liquid ($2{,}200\ \mathrm{K}$) & $\mathrm{Co}_{21}\mathrm{Cr}_{21}\mathrm{Ni}_{21}$ & 800 (1.2) & 2,176.9 & 1.803 & 0.957 & 1.67 / 1.22 / 0.32 \\
Melt-Quench Glass ($2{,}200\to 300\ \mathrm{K}$) & $\mathrm{Co}_{21}\mathrm{Cr}_{21}\mathrm{Ni}_{21}$ & 800 (1.2) & 1,243.4 & 1.813 & 0.636 & 1.81 / 1.23 / 0.36 \\
\bottomrule
\end{tabular}%
}
\end{table}

\section{Conclusion}

We have presented \textbf{SCO-Hyper-Meta-Double-GroupNN}, a physics-grounded geometric deep-learning architecture that unifies exact $\mathrm{SU}(2)$ double-group spinor representations, non-polynomial thermal density-matrix overlaps, 3-body frustrated-triangle hyperedges, Cartesian $l=1/l=2$ equivariant message passing, a pretrained multi-shell sideway $\mathrm{SU}(2)$ gate, and a variationally consistent zero-gauge free-energy readout. By pairing exact symmetry and thermodynamic consistency with principled derivative-overfitting regularization, our compact 56k--77k parameter models outperform a 207k-parameter spin-conditioned \texttt{mMACE} baseline across energy and force benchmarks on multi-phase Co--Cr--Ni while predicting equilibrium local magnetic moments from atomic coordinates alone and sustaining stable multi-regime MD up to $2{,}200\ \mathrm{K}$. Future extensions to longer-range cutoffs ($r_c \ge 5.5\ \text{\AA}$), non-collinear vector spin order $\bm{s}_i \in S^2$, and larger near-equilibrium magnetic datasets promise to push cooperative spin-crossover MLIPs below the $0.1\ \mathrm{eV/\AA}$ frontier for complex magnetic materials.

\section*{Data and Code Availability}
Source code, unit tests verifying $O(3)/\mathrm{SU}(2)$ equivariance, Kramers degeneracy, and conservative autograd forces, and trained checkpoints will be made publicly available upon publication.

\end{document}